\documentclass[aps,prb,twocolumn]{revtex4}
\usepackage{makeidx}
\usepackage{amsmath,amssymb,amsfonts,amsthm}
\usepackage{graphicx,bm}

\begin{document}

\title{Phase transitions and phase diagram of the ferroelectric perovskite
(Na$_{0.5}$Bi$_{0.5})_{1-x}$Ba$_{x}$TiO$_{3}$ by anelastic and dielectric
measurements}
\date{}
\author{F. Cordero,$^{1}$ F. Craciun,$^{1}$ F. Trequattrini,$^{2}$ E.
Mercadelli$^{3}$ and C. Galassi$^{3}$}
\affiliation{$^1$ CNR-ISC, Istituto dei Sistemi Complessi, Area della Ricerca di Roma -
Tor Vergata,\\
Via del Fosso del Cavaliere 100, I-00133 Roma, Italy}
\affiliation{$^{2}$ Dipartimento di Fisica, Universit\`{a} di Roma \textquotedblleft La
Sapienza\textquotedblright , P.le A. Moro 2, I-00185 Roma, Italy}
\affiliation{$^{3}$ CNR-ISTEC, Istituto di Scienza e Tecnologia dei Materiali Ceramici,
Via Granarolo 64, I-48018 Faenza, Italy}

\begin{abstract}
The complex elastic compliance and dielectric susceptibility of (Na$_{0.5}$Bi%
$_{0.5})_{1-x}$Ba$_{x}$TiO$_{3}$ (NBT-BT) have been measured in the
composition range between pure NBT and the morphotropic phase boundary
included, $0\leq x\leq 0.08$. The compliance of NBT presents sharp peaks at
the rhombohedral/tetragonal and tetragonal/cubic transitions, allowing the
determination of the tetragonal region of the phase diagram, up to now
impossible due to the strong lattice disorder and small distortions and
polarizations involved. In spite of ample evidence of disorder and
structural heterogeneity, the R-T\ transition remains sharp up to $x=0.06$,
whereas the T-C transition merges into the diffuse and relaxor-like
transition associated with broad maxima of the dielectric and elastic
susceptibilities. An attempt is made at relating the different features in
the anelastic and dielectric curves to different modes of octahedral
rotations and polar cation shifts. The possibility is also considered that
the cation displacements locally have monoclinic symmetry, as for PZT near
the morphotropic phase boundary.
\end{abstract}

\maketitle

% 77.80.B- Phase transitions and Curie point
% 77.84.Cg PZT ceramics and other titanates
% 77.22.Ch Permittivity (dielectric function)
% 62.40.+i Anelasticity

\section{Introduction}

The phase diagrams of the solid solutions of the ferroelectric perovskite Na$%
_{0.5}$Bi$_{0.5}$TiO$_{3}$ (NBT) with other tetrahedral perovskites are
actively studied, mainly with the intent of obtaining compositions close to
a morphotropic phase boundary (MPB) between rhombohedral (R) and tetragonal
(T)\ phases, with easily reorientable ferroelectric domains and high
electromechanical coupling. In fact, NBT based perovskites are among the
possible lead-free piezoelectric materials that would substitute the widely
used PbZr$_{1-x}$Ti$_{x}$O$_{3}$, when new environmental legislations will
limit the use of Pb containing materials.

The phase diagram of (Na$_{0.5}$Bi$_{0.5})_{1-x}$Ba$_{x}$TiO$_{3}$ (here
abbreviated as NBT-BT or BNBT-$100x$)\ is still uncertain in the interesting
region $x<0.1$ including the MPB, particularly regarding the nature and
extension of the presumably almost antiferroelectric (AFE) phase that
becomes ferroelectric (FE) on further cooling. Also the temperature
evolution of the structure of undoped NBT is the object of contradictory
interpretations.\cite{DTB08} The main difficulties in determining the
structure of NBT by neutron or x-ray diffraction are due to the very low
distortion in both the tetragonal and rhombohedral phases, with a cell
distortion from cubic of $<0.2\%$ and $\sim 0.4\%$ respectively,\cite{JT02}
and to the structural disorder in the Na/Bi sublattice, which is averaged
out and compensated by anomalously large thermal factors in the analysis.%
\cite{ASI09} In addition, x-ray diffraction is little sensitive to the
rotations of the O octahedra involved in the various structural
transformations, due to the low atomic number of O. Cation disorder is
enhanced in the solid solutions with other perovskites; it is responsible
for the diffuse nature of the lower temperature transitions in the
dielectric measurements, sometimes considered of the relaxor type, and
possibly for the very broad temperature range of coexistence between T and R
phases, with characteristics that vary from study to study. There is also
scarce reproducibility of the transition temperatures and of the associated
anomalies in the dielectric and other physical properties, which generally
exhibit very broad anomalies with large hysteresis between heating and
cooling.\cite{DTB08,FLG09} These uncertainties may also be connected with
differences from sample to sample in the local cation ordering and non
perfect stoichiometry in the Na/Bi sublattice, due to preferential loss of
Bi during sintering.

The elastic constants should provide clear information on the occurrence of
transformations between C, T and R phases, which are of ferroelastic nature,
but to our knowledge only few studies of the acoustic properties of NBT
exist,\cite{STS95,FLG09,Suc02} which have not an obvious connection with the
known phase transitions. Here we present an extensive study of (NBT)$_{1-x}$%
(BT)$_{x}$ with $0\leq x\leq 0.08$ by combined anelastic and dielectric
spectroscopies.

\section{Experimental}

Powders of the composition (Na$_{0.5}$Bi$_{0.5})_{1-x}$Ba$_{x}$TiO$_{3}$
with $x=0$ and $x=$ 2, 3, 4, 5, 6, 8~mol\% were prepared following the
mixed-oxide method. Stoichiometric amounts of Na$_{2}$CO$_{3}$ (Merck 6392),
Bi$_{2}$O$_{3}$ (Aldrich 223891), BaCO$_{3}$ (Merck 1714), TiO$_{2}$
(Degussa P 25) were mixed in ethanol with zirconia milling media for 48~h,
dried and sieved. For each composition the calcining and sintering
temperatures were systematically investigated in order to improve the cold
consolidation behavior and the final density. After heat treatment at
$700-800$~$^{\mathrm{o}}$C for 1~h and further milling and
drying, the powders were isostatically pressed at 300~MPa and sintered at
1150~$^{\mathrm{o}}$C for 2~h. In order to avoid the loss of Na and Bi,
which is significant at temperatures over 1000~$^{\mathrm{o}}$C, most of the
sintering processes were carried out with the samples placed on a ZrO$_{2}$
disc, covered with an Al$_{2}$O$_{3}$ crucible and sealed with NBT pack. The
x-ray diffraction analysis evidenced the pure perovskitic phase in all the
samples investigated. The microstructure of the sintered samples was
observed on polished and etched surfaces by means of scanning electron
microscopy (Leica Cambridge Stereoscan 360). Depending on the composition
(at the same sintering conditions) the microstructures result more or less
homogeneous, with some large, squared grains in a matrix of smaller grains;
among those sintered in the NBT pack powder, pure NBT showed the largest
grain size ($5-20$~$\mu$m), which reduced upon addition of barium, the lowest
mean size being found with $x=0.08$ (0.5~$\mu$m).

We prepared discs (diameter 22~mm, thickness 2~mm) for the dielectric
measurements and blocks ($60\times 7\times 7$~mm$^{3}$) subsequently cut
into bars  ($40\times 4\times 0.6$~mm$^{3}$) for both anelastic and
dielectric experiments. The faces of the discs were ground, silver
electroded and poled with an electric field of 3~kV/mm in a
silicon oil bath at 120~$^{\mathrm{o}}$C for 30 min, while the bars were
tested unpoled.

The dielectric measurements were carried out on poled and unpoled discs and
bars in a frequency range between 200~Hz and 200~kHz using a Hewlett-Packard
LCR meter (HP 4284A) with a four wire probe and an ac driving signal level
of 0.5~V/mm. The dielectric permittivity $\varepsilon =\varepsilon ^{\prime
}-i\varepsilon ^{\prime \prime }$ was obtained from the measured values of
capacitance and loss $\tan \delta =\varepsilon ^{\prime \prime }/\varepsilon
^{\prime }$. The measurements were made on heating/cooling at $1-1.5$~K/min
between 300 and 570~K in a Delta Design climatic chamber model 9023A.

The complex Young's modulus $E=E^{\prime }+iE^{\prime \prime }$ was measured
by electrostatically exciting the flexural vibrations of the bars, which
were suspended on thin thermocouple wires in vacuum. It was possible to
excite the 1st, 3rd and sometimes 5th flexural modes, whose resonating
frequencies are in the ratios $1:5.4:13$. The fundamental frequency is given
by\cite{NB72}
\begin{equation*}
f=1.028\frac{h}{l^{2}}\sqrt{\frac{E}{\rho }}
\end{equation*}%
where $l$ is the length, $h$ the thickness, $\rho $ the density the bar; for
most samples it was $f\sim 1.4$~kHz. From the above formula it is possible
to extract the absolute value of the Young's modulus. Such a value is
affected by an error that may easily be of the order of 10\%, due to
uncertainties in the sample dimensions, porosity (the density was always $%
93.7-98.6\%$ of the theoretical one) and above all due to the Ag electrode,
which was applied to one or two faces and whose thickness and Young's
modulus could not be controlled. The anelastic spectra are displayed in
terms of the real part of the reciprocal Young's modulus or compliance, $s=$
$E^{-1}=$ $s^{\prime }-is^{\prime \prime }$, and of the elastic energy loss
coefficient $Q^{-1}=$ $s^{\prime \prime }/s^{\prime }$. Since the anelastic
spectroscopy experiments were made in vacuum, some oxygen loss occurred when
800~K were exceeded, and the sample became darker. In most cases, however,
we did not note any effect on the anelastic and even dielectric spectra;
major changes occurred if 1000\ K were exceeded with samples prepared
without precautions against Bi loss during sintering.

\section{Results and Discussion}

\subsection{Undoped NBT}

\subsubsection{The anelastic spectrum of NBT}

\begin{figure}[tbp]
\includegraphics[width=8.5 cm]{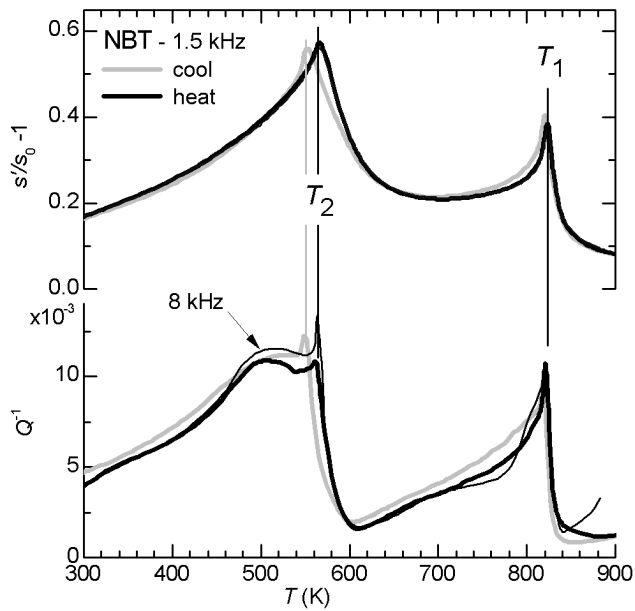}
\vspace{-4.5 cm}
\caption{Reciprocal Young's modulus $%
s^{\prime }=1/E^{\prime }$ and elastic energy loss $Q^{-1}$ of NBT measured
during heating (black) and cooling (gray) at 1.4~kHz (thick lines) and
7.4~kHz (thin line, only heating).}
\label{fig1}
\end{figure}

Figure \ref{fig1} shows the compliance curves
$s^{\prime }\left( T\right) $ and the losses $Q^{-1}\left( T\right) $ of
undoped NBT measured on heating and cooling at 1.5~kHz (thick black and grey
lines); the thin black curves were measured at 8~kHz during heating and are
practically coincident with those at 1.5~kHz, especially in the real part.
This means that the anomalies in $s^{\prime }$ are not due to relaxation of
domain walls or defects, but they reflect the intrinsic elastic constants
plus the fast relaxation of the order parameters involved in the structural
transformations, which can instantaneously follow the sample vibration. Some
frequency dependence is observed in the losses, and may be due to the
relaxation of domain walls. There are sharp peaks at $T_{1}=823$~K and $%
T_{2}=564$~K (during heating) in the complex compliance, whose sharpness is
in contrast with the broad anomalies in the other physical properties,
including the acoustic ones in the GHz range, and with the structural
changes, which are faint and spread over wide temperature ranges. The
anomalies appear at slightly lower temperature during cooling, with
hystereses of 3 and 13~K for $T_{1}$ and $T_{2}$, respectively.

% \FRAME{ftbpFU}{3.3909in}{3.2188in}{0pt}{\Qcb{Reciprocal Young's modulus $%
% s^{\prime }=1/E^{\prime }$ and elastic energy loss $Q^{-1}$ of NBT measured
% during heating (black)\ and cooling (gray) at 1.4~kHz (thick lines)\ and
% 7.4~kHz (thin line, only heating).}}{\Qlb{fig1}}{fig1.tif}{\special{language
% "Scientific Word";type "GRAPHIC";maintain-aspect-ratio TRUE;display
% "USEDEF";valid_file "F";width 3.3909in;height 3.2188in;depth
% 0pt;original-width 3.3442in;original-height 3.1739in;cropleft "0";croptop
% "1";cropright "1";cropbottom "0";filename 'fig1.tif';file-properties
% "XNPEU";}}

\subsubsection{Phase transitions in NBT}

The picture of the structural transformations in pure NBT is still confused
and contradictory; therefore, in order to relate the elastic anomalies of
Fig. \ref{fig1} to structural changes, we replotted our data in Fig. \ref%
{fig ov}a), together with the dielectric susceptibility and the
temperature dependencies of various structural parameters measured
in other experiments.

% \FRAME{ftbpFU}{8.6042cm}{12.7228cm}{0pt}{\Qcb{Temperature depenndece of
% various properties of NBT from our measurements and from the literature, as
% explained in the text.}}{\Qlb{fig ov}}{overview.tif}{\special{language
% "Scientific Word";type "GRAPHIC";maintain-aspect-ratio TRUE;display
% "USEDEF";valid_file "F";width 8.6042cm;height 12.7228cm;depth
% 0pt;original-width 5.9127in;original-height 8.745in;cropleft "0";croptop
% "1";cropright "1";cropbottom "0";filename 'overview.tif';file-properties
% "XNPEU";}}

\begin{figure}[tbp]
\includegraphics[width=8.5 cm]{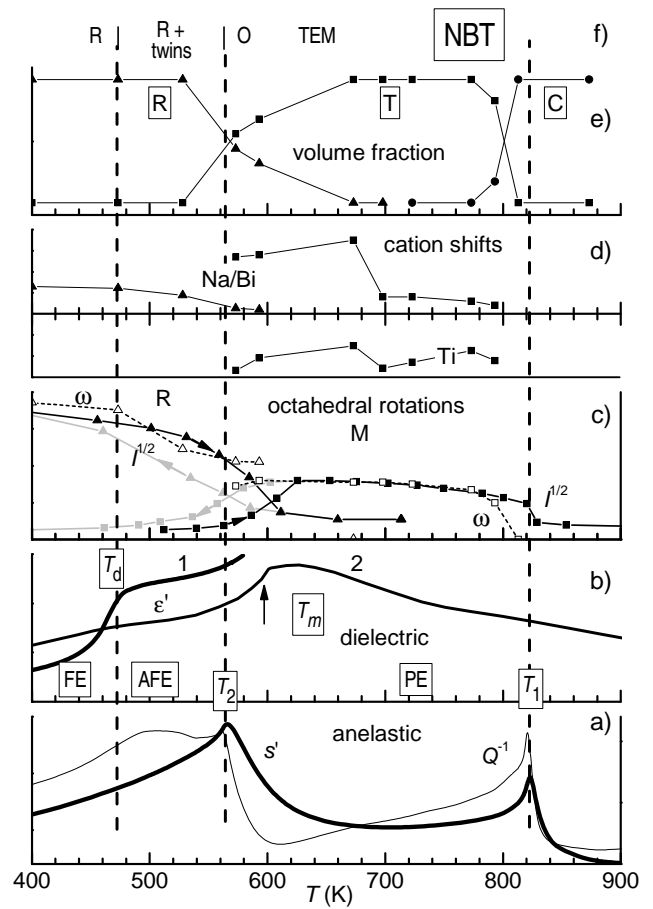}
\caption{Temperature depenndece of
various properties of NBT from our measurements and from the literature, as
explained in the text.}
\label{fig ov}
\end{figure}

For clarity, the labels of the ordinates are omitted; only
the temperature dependences are relevant here. Starting from the bottom
panel a), the curves $s^{\prime }$ and $Q^{-1}$ measured during heating
determine the temperatures $T_{1}$ and $T_{2}$. In the next panel
(ordinates: $0\leq \varepsilon ^{\prime }\leq 3000$) curve 1 is the
dielectric permittivity $\varepsilon ^{\prime }$ measured during heating on
poled NBT at 10~kHz. The step at $\sim 470$~K occurs at the same temperature
where the macroscopic polarization vanishes, and the $P-E$ hysteresis loops
become narrow or pinched in the center, and which is therefore called
depolarization temperature $T_{d}$; it defines the transition between
ferroelectric (FE) and possibly antiferroelectric (AFE)\cite%
{TSS94,OK06,HNT08} state. The antiferroelectric nature of the phase at $%
T>T_{d}$ is not unanimously accepted\cite{VIK85,Suc98,BKN06} and is deduced,
besides from the shape of the $P-E$ hysteresis curves, from the broad
maximum of $\varepsilon ^{\prime }$ around $T_{m}\sim 630$~K, which would
signal the transition from AFE to paraelectric (PE). The width of this
maximum, a possible slight frequency dispersion, especially at high
frequencies,\cite{PKF04} the fact that no structural transformation at $%
T_{m} $ but instead several indications of phase coexistence on nanoscale
exist,\cite{AIS06,DTB09,ASI09} induce several authors to liken this diffuse
transition to a relaxor transition not only in the solid solutions with BT
or other ferroelectrics, but also in pure NBT.\cite%
{LGS00,KBD03,HPK06,TWL07,FLG09,ASI09} The frequency dispersion is however
definitely smaller than for traditional relaxors.\cite{TSS94,PKF04} Due to
limitations in the temperature range of our apparatus, we could neither
reach the T-C\ transition nor the maximum of $\varepsilon ^{\prime }$ at $%
T_{m}$, and therefore we report curve 2 from Ref. \onlinecite{OK06}, with a
reduced step at $T_{d}$ and an additional step, indicated by an arrow,
connected with the R-T transition at a temperature higher than our $T_{2}$.
The position, amplitude and shape of such a step is highly variable from
experiment to experiment,\cite{TSS94,HNT06} while a dielectric anomaly has
never been observed at $T_{1}$.

Figure \ref{fig ov}c) presents the temperature dependence of the rotations
of the octahedra determined by neutron diffraction\cite{VIK85,JT02}
(ordinates: $0\leq \omega \leq 8^{\mathrm{o}}$). In the case of NBT, the
tetragonal structure arises from rotations of the octahedra about their $c$
axes ($a^{0}a^{0}c^{+}$ in Glazer's notation\cite{Gla72}), giving rise to
superlattice peaks from the M point of the Brillouin zone, while the
rhombohedral structure is due to rotations about all three principal
directions of the cubic cell, with alternate sign along each axis ($%
a^{-}a^{-}a^{-}$ in Glazer's notation), giving rise to R-type superlattice
peaks. The closed symbols (black = heating and gray = cooling) are the
square root of the intensities of these superlattice reflections,\cite{VIK85}
which should be proportional to the squares of the angles $\omega $ of
rotations of type M and R.\cite{NC87} They are scaled in order to overlap
with $\omega $ measured from Rietveld refinements\cite{JT02} (open symbols).
The curves from the two experiments\cite{VIK85,JT02} are in reasonable
agreement with each other and represent the rotational order parameters of
the two transitions. On cooling from the C phase, the M rotation (squares)
has an onset at $T_{1}\sim 820\pm 5$~K, no hysteresis between heating and
cooling and a precursor tail extending a hundred kelvin above $T_{1}$, at
least in the experiment of Vakhrushev \textit{et al.}\cite{VIK85} This
behavior is close to what expected for the condensation of a soft M phonon
mode at the zone boundary, acting as order parameter of the T phase. The
softening of such a mode has been deduced from the intensity of the
quasielastic neutron scattering at the M point,\cite{VIK85} but not observed
by other techniques, to our knowledge. On further cooling, the R rotations
(triangles) start being observed below $\sim 610$~K and the amplitude of the
M rotations decreases. On the basis of these curves, the onset temperature
for the R-T transformation cannot be defined with an uncertainty smaller
than 40~K and there is a hysteresis of 60~K between heating and cooling.
Again, a softening of the R phonon mode in the T phase has been deduced from
the intensity of the quasielastic scattering at the R point.\cite{VIK85}

The cation shifts with respect to the octahedra deduced from the neutron
Rietveld refinements\cite{JT02} are plotted in Fig. \ref{fig ov}d). In the
ferroelectric R phase both the Na/Bi and Ti (not shown) shifts are parallel
to the pseudocubic $\left[ 111\right] $ direction; in the T phase, they are
parallel to the tetragonal axis but with opposite signs, so that the
structure is only weakly polar, in agreement with the nearly
antiferroelectric properties in this temperature region. It is not clear
whether the jump in cation shifts near 680~K may be put in relation with the
broad dielectric maximum at $T_{m}$. The refinements of the neutron
scattering data of Jones and Thomas\cite{JT02} indicate
regions of coexistence of C and T phases below $T_{1}$ and R and T\ phases
around $T_{2}$ (volume fractions in Fig. \ref{fig ov}e)).
Unfortunately, the large temperature intervals between
successive measurements did not allow to clarify the controversy on the R-T
coexistence regions, as discussed by the same authors.\cite{JT02} A later
diffuse neutron scattering study\cite{BKN06} indicates a non trivial
transformation between R and T\ structures over an extremely broad range of
temperatures with a residual modulation of the R phase along the $c$ axis of
the parent structure. A modulated R structure is found also in recent TEM\
experiments,\cite{DTB08,DTB09} which are interpreted in terms of a low
temperature $R3c$ structure, which above 473~K starts twinning parallel to
the $\left[ 010\right] $ pseudocubic planes. On further heating, the density
of $Pnma$ twin planes increases and the modulation of the polarization
direction gives rise to antiferroelectric properties until at $563$~K the
modulation disappears and the structure becomes homogeneous, identified as $%
Pnma$ orthorhombic rather than tetragonal.\cite{DTB09} The temperatures of
appearance and disappearance of the striation in the TEM images are
indicated as vertical bars in Fig. \ref{fig ov}e), and correspond to $T_{d}$
and $T_{2}$.

\subsubsection{Identification of the elastic anomalies}

Based on Fig. \ref{fig ov} and the above discussion we can identify the
temperatures $T_{1}$ and $T_{2}$ of the peaks in the elastic compliance with
the onsets of the T-C and R-T transformations. The first case is obvious, in
view of the coincidence of $T_{1}$ with the onset of the M-type rotations of
the octahedra (Fig. \ref{fig ov}c) causing the tetragonal distortion. The
fact that the T-C transformation is evident in the elastic but absent in the
dielectric susceptibility\cite{TSS94,OK06} confirms that it is triggered by
the non-polar rotational mode of the octahedra, as already discussed in
connection with optical and high-frequency dielectric measurements.\cite%
{PKF04} The small hysteresis of 3~K between heating and cooling indicates
that the transition is almost 2nd order. According to the Landau theory of
phase transitions without fluctuations, the expected effect on the elastic
constants involved in the resulting tetragonal distortion is a negative step
when passing from cubic to tetragonal at $T_{1}$,\cite{CS98} or equivalently
a positive step in the compliances. The elastic anomaly at $T_{1}$ has a
consistent peaked component besides the step, but clearly corresponds to the
almost 2nd order T-C transition. The occurrence of the additional peaked
softening is common in perovskites and is usually explained in terms of
fluctuation effects;\cite{CS98} it will not be discussed further here.

% \FRAME{ftbpFU}{8.6042cm}{10.01cm}{0pt}{\Qcb{Left ordinates, continuous
% lines: dielectric permittivity $\protect\varepsilon ^{\prime }$ and losses $%
% \tan \protect\delta $ of NBT measured at 1~kHz on heating (thick) and
% cooling (thin). Right ordinates, circles: elastic compliance $s^{\prime }$
% and energy loss $Q^{-1}$ of NBT measured at 1.5~kHz on heating (closed) and
% cooling (open).}}{\Qlb{fig NBT}}{nbt.tif}{\special{language "Scientific
% Word";type "GRAPHIC";maintain-aspect-ratio TRUE;display "USEDEF";valid_file
% "F";width 8.6042cm;height 10.01cm;depth 0pt;original-width
% 2.879in;original-height 3.6858in;cropleft "0";croptop "1";cropright
% "1";cropbottom "0";filename 'NBT.tif';file-properties "XNPEU";}}

\begin{figure}[tbp]
\includegraphics[width=8.5 cm]{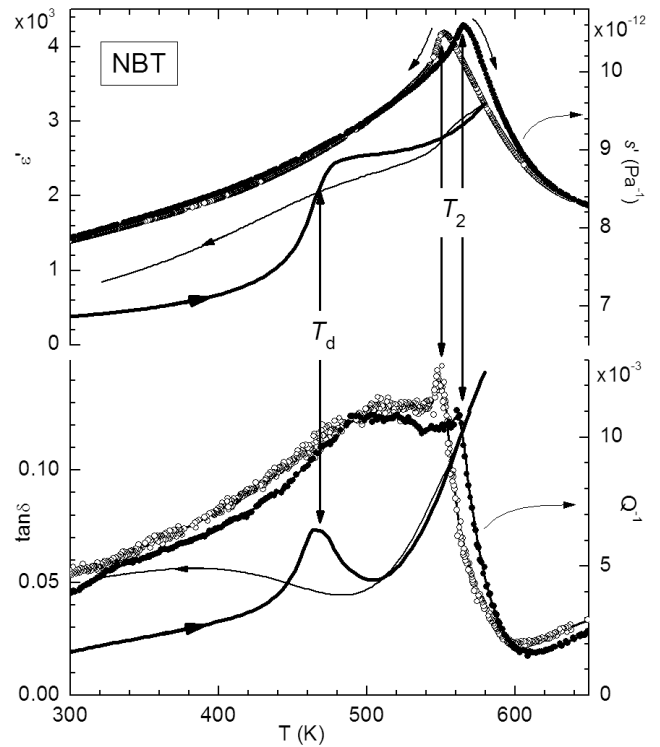}
\vspace{-2.5 cm}
\caption{Left ordinates, continuous
lines: dielectric permittivity $\protect\varepsilon ^{\prime }$ and losses $%
\tan \protect\delta $ of NBT measured at 1~kHz on heating (thick) and
cooling (thin). Right ordinates, circles: elastic compliance $s^{\prime }$
and energy loss $Q^{-1}$ of NBT measured at 1.5~kHz on heating (closed) and
cooling (open).}
\label{fig NBT}
\end{figure}

The R-T\ transition is characterized by a variety of apparently
contradictory experimental results, which generally agree on the diffuseness
of the transition and broad range of coexistence of the two phases; the
competition between the two M\ and R rotational instabilities even suggests
the possibility of glassy behavior.\cite{BKN06} In some dielectric
measurements there is a step near the region of R-T coexistence, but only on
heating and not on cooling\cite{TSS94,Suc01} and with varying temperatures
and shapes.\cite{Suc01,QSS05,OK06} The dielectric permittivity $\varepsilon
^{\prime }$ and losses $\tan \delta $ of one of our samples are plotted in
Fig. \ref{fig NBT} as thick (heating) and thin (cooling) lines, together
with the compliance $s^{\prime }$ and elastic losses $Q^{-1}$ as closed
(heating) and open (cooling)\ symbols. In spite of the limited maximum
temperature, it is possible to see a step in the $\varepsilon ^{\prime }$
curves that correlates very well with the peak in $s^{\prime }$. These
compliance curves $s^{\prime }$ and particularly the acoustic losses are
sharply peaked at a temperature $T_{2}$ with only 13~K of hysteresis between
heating and cooling and allow us to define a precise temperature for the R-T
transition (note the sharpness of the losses at higher frequency at $T_{2}$
in Fig. \ref{fig1}). In the light of the most recent experiments,\cite%
{BKN06,DTB08,DTB09} the temperature $T_{2}$ can be identified with the onset
of the appearance of the R\ structure, which however appears gradually as a
modulation between the M-type in-phase tilts of the octahedra about the $c$
axis in the T phase and the R-type antiphase tilts of the R phase,\cite%
{BKN06} or as finely twinned R phase.\cite{DTB09} Again, the fact that the
anomaly is sharp and pronounced in the elastic compliance and less defined
in the dielectric susceptibility suggests that the primary order parameter
is the R-type tilting of the octahedra, which in turn induces a change in
the direction and amplitude of the cation shifts. Conversely, there is no
trace of the ferroelectric transition at $T_{d}$ in the elastic compliance.
This may also be due to the fact that the ferroelectric transition does not
seem to involve a change in the structure, but only in the size and
correlation length of the domains.

The present measurements do not allow yet to unify the different pictures of
the R phase obtained from various techniques into a coherent one, but
indicate that, in spite of the various indications of extreme heterogeneity
on the local scale, short correlation lengths and relaxor-like behavior, the
R-type instability responsible for the R phase produces an elastic anomaly
comparable to those of other crystalline perovskites with normal structural
transitions occurring at a well defined temperature.

The above data can be compared with $c_{11}\left( T\right) $ measured on a
NBT crystal with the ultrasonic pulse echo technique,\cite{Suc02} where a
shallow minimum is found near 580~K, with $\sim 25$~K of hysteresis between
heating and cooling.\emph{\ }The minimum has been attributed to relaxation
of the polar regions rather that to the phase transition itself, and,
although its temperature is close to $T_{2}$, it is much shallower than the
peak in the reciprocal Young's modulus $s^{\prime }\left( T\right) $.
Considering that the phase transition at $T_{2}$ involves the rhombohedral
distortion, it is possible that the sharper response of the
Young's modulus is mainly due to the shear elastic constants of $c_{44}$\
symmetry. Two Brillouin scattering experiments, which probe the acoustic
properties at much higher frequencies, yield still different results. In one
case\cite{STS95} a dip in sound velocity and a peak in damping are found
near 700~K, which, instead of coinciding with one of the temperatures of the
T-C and R-T\ transitions, is between the two. The observation has been
explained in terms of strong fluctuations of the two coupled order
parameters for the two T and R\ phases, plus disorder in the Na/Bi
sublattice and has been put in parallel to relaxor ferroelectrics. In a more
recent experiment,\cite{FLG09} the intensity of the quasi-elastic light
scattering and a distortion of the phonon peaks from a lorentzian shape are
found to be peaked at 550~K, which is about $T_{2}$.

\subsection{Samples quality}

It is true that different techniques may produce different pictures
of a same system; for example local probes like the NMR or EXAFS
spectroscopies indicate larger and more disordered cation
displacements than diffraction techniques.\cite{ASI08} Yet, the
quality of samples is important for NBT, since different samples may
give different results from the same technique, notably in the
dielectric and ferroelectric properties. The main issues are cation
ordering, stoichiometry and related defects providing ionized
electronic charges. Regarding cation ordering, there are
ferroelectric perovskites like Pb(Sc$_{0.5}$Ta$_{0.5}$)O$_{3}$
(PST), where the level of Sc/Ta ordering can be controlled and
produces marked effects on the temperature and diffuseness of the
ferroelectric transition.\cite{SC80} In NBT, in some case
superlattice $\frac{a}{2}\left( 111\right) $ peaks have been
observed by x-ray diffraction \cite{PCK94} or TEM\cite{LYH01,LHK02}
and attributed to an alternation of Bi and Na along the three
directions, but usually cation order is found to be very weak and
localized\cite{WR05} or absent,\cite{JT02} so that a random
distribution is assumed.\cite{PKF04} The level of cation ordering in
NBT and its solid solutions with other perovskites is not
controllable as in PST, and prolonged annealings in order to improve
ordering might rather cause loss of Bi,\cite{HNT09} so that a spread
of the NBT properties from sample to sample is not surprising. It is
therefore
appropriate to check whether the sharp elastic anomalies of NBT in Fig. \ref%
{fig1} are the result of a sample of exceptional quality.

\begin{figure}[tbp]
\includegraphics[width=8.5 cm]{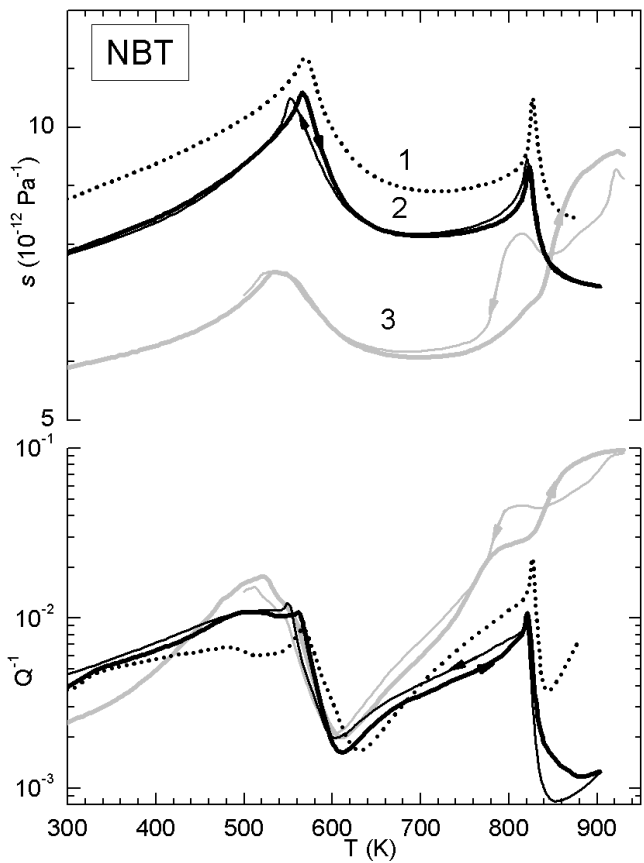}
\vspace{-1 cm}
\caption{Reciprocal Young's modulus $s^{\prime }$ and
elastic energy loss $Q^{-1}$ measured on heating (thick lines) and
cooling (thin lines) at $\sim 1.4$~kHz on three samples of NBT
prepared as described in the text.} \label{fig NBTsamples}
\end{figure}

Figure \ref{fig NBTsamples} shows the anelastic spectra of three samples of
NBT prepared in different ways. Samples 1 and 2 were calcined for 1~h at 800~%
$^{\mathrm{o}}$C (sample 1) or 700~$^{\mathrm{o}}$C (sample 2) and sintered
for 2 ~h at 1150~$^{\mathrm{o}}$C with surrounding NBT packs in order to
avoid Bi loss, while sample 3 was prepared with the same temperature
schedules of sample 1 but without precautions against Bi loss. Sample 2 is
the one of Fig. \ref{fig1}.

% \FRAME{ftbpFU}{3.3658in}{4.5221in}{0pt}{\Qcb{Reciprocal Young's modulus $%
% s^{\prime }$ and elastic energy loss $Q^{-1}$ measured on heating (thick
% lines) and cooling (thin lines) at $\sim 1.4$~kHz on three samples of NBT
% prepared as described in the text.}}{\Qlb{fig NBTsamples}}{nbtsamples.tif}{%
% \special{language "Scientific Word";type "GRAPHIC";maintain-aspect-ratio
% TRUE;display "USEDEF";valid_file "F";width 3.3658in;height 4.5221in;depth
% 0pt;original-width 3.32in;original-height 4.4693in;cropleft "0";croptop
% "1";cropright "1";cropbottom "0";filename 'NBTsamples.tif';file-properties
% "XNPEU";}}
% % expdata\FE\NBT\BNBT0all.OPJ

It is apparent that the Bi/Na stoichiometry is an issue, since in sample 3
only a reminiscence of the T-C transition at $T_{1}$ remains with the sign
of the step in the compliance even reversed, and also the transition at $T_{2}$
is smeared; in addition, strong irreversible relaxation phenomena occur
above 750~K, which are absent in the stoichiometric samples. Such relaxation
processes appear as an elastic energy loss two orders of magnitude larger
than in the cubic phase of stoichiometric samples, and are presumably due to
the migration of the light Na ion by a vacancy mechanism in the cation
deficient Na/Bi sublattice. The effect is absent in a perfectly
stoichiometric lattice. The dielectric susceptibilities of two samples
prepared as sample 1 and 3 (not shown here) demonstrate that the less
stoichiometric samples of type 3 are indeed of extremely bad quality also
from the dielectric point of view, with enormous losses due to mobile
charges, which totally conceal the transition at $T_{d}$. On the other hand,
samples of type 1 present $\varepsilon \left( T\right) $ curves entirely
comparable to those of "good" samples in the literature; therefore, the
sharp elastic anomalies at $T_{2}$ should not be ascribed to an exceptional
quality of those samples, but are a general property of NBT. In addition,
the elastic anomaly at $T_{2}$ clearly persists, although broadened, also in
less stoichiometric samples of type 3, where the other transitions are
washed out. This demonstrates that the rotational instability leading to the
R-T transformation is also robust against lattice disorder.

\subsection{Anelastic spectra of NBT-BT}

\begin{figure}[tbp]
\includegraphics[width=8.5 cm]{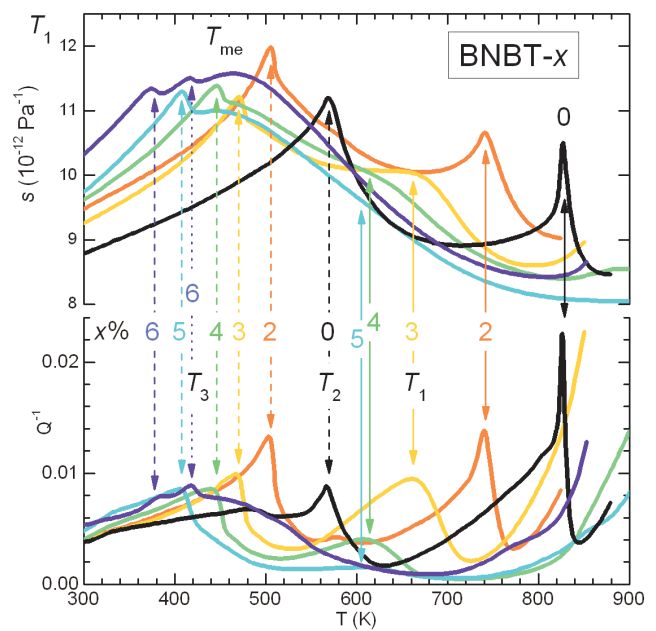}
\vspace{-4 cm}
\caption{Reciprocal Young's modulus $%
s^{\prime }$ and elastic energy loss $Q^{-1}$ of a series of (NBT)$_{1-x}$%
(BT)$_{x}$ samples with $0\leq x\leq 0.06$ measured at $\sim 1.4$~kHz. The
solid arrows indicate $T_{1}$, the dashed arrows $T_{2}$ and the dotted $%
T_{3}$ (Color online).}
\label{fig anelvsx}
\end{figure}

Figure \ref{fig anelvsx} presents compliance $s^{\prime }$ and elastic
energy losses $Q^{-1}$ curves measured on heating of a series of samples
with Ba substitution up to 6\%. We do not include the curves for 8\% for
clarity and because we consider those measurements preliminary, due to the
sample quality.

% \FRAME{ftbpFU}{3.3909in}{3.3062in}{0pt}{\Qcb{Reciprocal Young's modulus $%
% s^{\prime }$ and elastic energy loss $Q^{-1}$ of a series of (NBT)$_{1-x}$%
% (BT)$_{x}$ samples with $0\leq x\leq 0.06$ measured at $\sim 1.4$~kHz. The
% solid arrows indicate $T_{1}$, the dashed arrows $T_{2}$ and the dotted $%
% T_{3}$ (Color online).}}{\Qlb{fig anelvsx}}{anelvsx.tif}{\special{language
% "Scientific Word";type "GRAPHIC";maintain-aspect-ratio TRUE;display
% "USEDEF";valid_file "F";width 3.3909in;height 3.3062in;depth
% 0pt;original-width 3.3442in;original-height 3.2612in;cropleft "0";croptop
% "1";cropright "1";cropbottom "0";filename 'anelvsx.tif';file-properties
% "XNPEU";}}

Upon substitution of Ba, the T-C\ transition lowers its temperature $T_{1}$
(continuous arrows), broadens and fades away, until at $x=0.06$ it is no
more discernible from the broad maximum that develops around $T_{me}\sim 450$%
~K. The R-T transition also lowers its temperature $T_{2}$ (dashed arrows),
and the corresponding elastic anomaly looses intensity and sharpness, but
remains always clearly visible. The peak in both the real and imaginary
parts of the compliance maintains a rather sharp component at a decreasing $%
T_{2}$, but its weight decreases in favor of a broad maximum at $T_{me}$
which is predominant already at $x=0.03$. The existence of two distinct
anomalies at $T_{2}$ and $T_{me}$ can be related to fact that there are at
least three interacting but distinct order parameters, namely M-type and
R-type rotations and cation shifts. Accordingly, the compliance presents
three main anomalies: the one at $T_{1}$ is obviously related to the M
rotations responsible for the tetragonal distortion, while the other two
should involve R-type rotations and cation displacements. The identification
of the step with the R/T\ transition, and therefore the R-type rotations, is
clear at $x=0$ and $x=0.06$, where the temperature $T_{2}$ so defined
coincides with that determined for the R/T transition by other techniques,
and should therefore hold also for the intermediate temperatures (see also
the phase diagram below). We remain with the cation shifts as the main
responsible for the maximum at $T_{me}$.

\subsubsection{The anomaly at $T_{3}$}

% \FRAME{ftbpFU}{8.58cm}{9.9946cm}{0pt}{\Qcb{Left ordinates, continuous lines:
% dielectric permittivity $\protect\varepsilon ^{\prime }$ and losses $\tan
% \protect\delta $ of a sample of (NBT)$_{1-x}$(BT)$_{x}$ with $x=0.06$
% measured at 1~kHz on heating (thick) and cooling (thin). Right ordinates,
% circles: elastic compliance $s^{\prime }$ and energy loss $Q^{-1}$ measured
% at 1.4~kHz on heating; dashed lines: repetition of the measurement (heating
% and subsequent cooling) after one week of aging at room temperature. }}{\Qlb{%
% fig BNBT6}}{bnbt6.tif}{\special{language "Scientific Word";type
% "GRAPHIC";maintain-aspect-ratio TRUE;display "USEDEF";valid_file "F";width
% 8.58cm;height 9.9946cm;depth 0pt;original-width 3.3442in;original-height
% 2.6662in;cropleft "0";croptop "1";cropright "1";cropbottom "0";filename
% 'BNBT6.tif';file-properties "XNPEU";}}

\begin{figure}[tbp]
\includegraphics[width=8.5 cm]{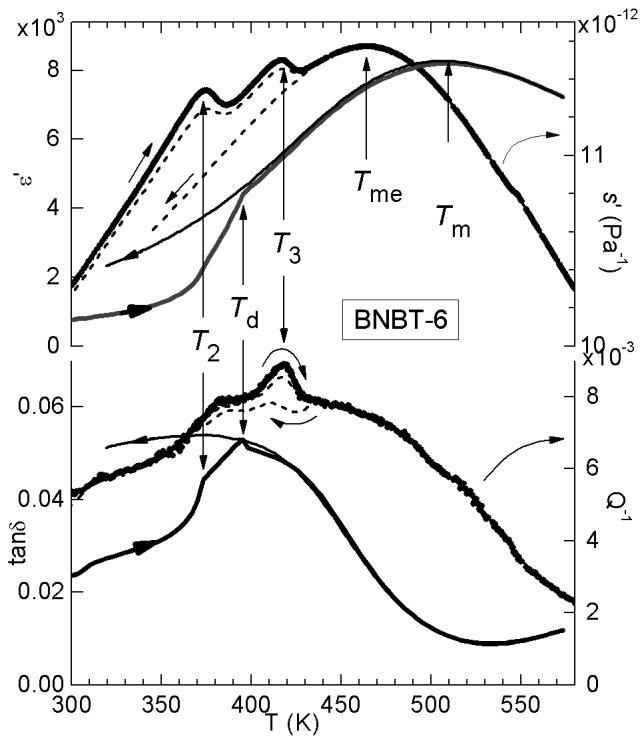}
\vspace{-2.5 cm}
\caption{Left ordinates, continuous lines:
dielectric permittivity $\protect\varepsilon ^{\prime }$ and losses
$\tan \protect\delta $ of a sample of (NBT)$_{1-x}$(BT)$_{x}$ with
$x=0.06$ measured at 1~kHz on heating (thick) and cooling (thin).
Right ordinates, circles: elastic compliance $s^{\prime }$ and
energy loss $Q^{-1}$ measured at 1.4~kHz on heating; dashed lines:
repetition of the measurement (heating and subsequent cooling) after
one week of aging at room temperature.} \label{fig BNBT6}
\end{figure}

At $x=0.06$ a new anomaly appears at a temperature indicated as $T_{3}$
(dotted arrow)\ in Fig. \ref{fig anelvsx}. It would be natural to identify
it with the ferroelectric transition at $T_{d}$, but two observations cast
doubts on this assignment. One is that $T_{d}$ deduced from the kink in $%
\varepsilon ^{\prime }$ and from the separation of the dielectric curves measured
during heating and cooling is slightly different from $T_{3}$. Figure \ref%
{fig BNBT6} compares the anelastic and dielectric curves of two samples of
BNBT-6, where the correlation between the anomalies at $T_{2}$ is excellent
(hardly visible in $\varepsilon ^{\prime }$ but clear in $\tan \delta $) but
the identification of $T_{3}$ with $T_{d}$ is dubious. The other observation
is that in the elastic compliance of NBT\ there is no trace of $T_{d}$ (see
Fig. \ref{fig1}) and one would expect the same behavior at $x=0.06$, unless
the ferroelectric transition has changed nature. Indeed, at $x=0$ the
temperature $T_{d}$ separates a FE-R from a AFE-R phase, while at $x=0.06$
it separates the FE-T phase from a presumably AFE-C or pseudocubic one.
Notice that the anomalies of both permittivity and compliance are evident
during heating but disappear during cooling, indicating that in the highly
disordered NBT-BT there is a slow structural evolution in the R phase,
presumably domain coarsening. This is confirmed by the fact that the anomalies on
heating require few days of aging at room temperature in order to fully
develop. The dashed anelastic curves in Fig. \ref{fig BNBT6} are the result
of one week of aging at room temperature after a measurement at high
temperature, and the two elastic anomalies are not yet fully developed.

\subsection{Dielectric spectra}

The dielectric spectra have been extensively discussed,\cite%
{HWN07,Isu05b,SM01,GSM04,LYH04b,ZLL09} and here we point out only the main
conclusions and some differences with respect to the previous literature.

% \FRAME{ftbpFU}{8.6042cm}{8.4175cm}{0pt}{\Qcb{Real and imaginary parts of the
% dielectric permittivity of a series of samples with $0\leq x\leq 0.08$
% measured during heating at 10~kHz.}}{\Qlb{fig diel}}{dielall.tif}{\special%
% {language "Scientific Word";type "GRAPHIC";maintain-aspect-ratio
% TRUE;display "USEDEF";valid_file "F";width 8.6042cm;height 8.4175cm;depth
% 0pt;original-width 5.9196in;original-height 5.7917in;cropleft "0";croptop
% "1";cropright "1";cropbottom "0";filename 'dielAll.tif';file-properties
% "XNPEU";}}

\begin{figure}[tbp]
\includegraphics[width=8.5 cm]{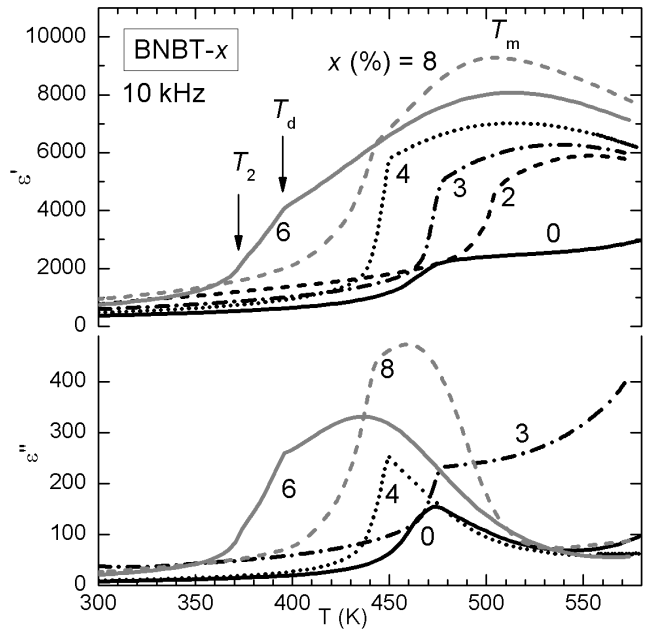}
\vspace{-4 cm} \caption{Real and imaginary parts of the dielectric
permittivity of a series of samples with $0\leq x\leq 0.08$ measured
during heating at 10~kHz.} \label{fig diel}
\end{figure}

In Fig. \ref{fig diel} a panoramic of the dielectric permittivity and losses
measured at 10~kHz during heating of (NBT)$_{1-x}$(BT)$_{x}$ with $0\leq
x\leq 0.08$ is shown. The samples with 2 and 3\% Ba had rather high
conductivities and hence losses, but the real part $\varepsilon ^{\prime }$ at
10~kHz, where the influence of such conductivity effects is reduced, fit
well the general trend of growing maximum of $\varepsilon ^{\prime }$ at $%
T_{m}$ and lowering $T_{d}$ with increasing $x$. The step at $T_{2}$ is
invisible in this scale at $x=0$ (but see Fig. \ref{fig NBT}),\ it is
indistinguishable from the ferroelectric transition due to the coincidence
of $T_{2}$ and $T_{d}$ for $x\geq 0.02$ but it is again separated from $%
T_{d} $ at $x=0.06$. A first difference between our measurements and those
of Hiruma \textit{et al.}\cite{HWN07} is that we see a single dielectric
anomaly already at $x=0.02$ while they observe such a merging at $x=0.03$.
We also observed that temperature and width of these steps depend on the
sample preparation and the polarization status, being generally sharper and
at higher temperature in poled samples.

% \FRAME{ftbpFU}{8.6042cm}{8.9469cm}{0pt}{\Qcb{Real and imaginary parts of the
% permittivity $\protect\varepsilon $ of an initially poled sample of (NBT)$%
% _{1-x}$(BT)$_{x}$ with $x=0.04$ (Color online).}}{\Qlb{fig BNBT4diel}}{%
% bnbt4diel.tif}{\special{language "Scientific Word";type
% "GRAPHIC";maintain-aspect-ratio TRUE;display "USEDEF";valid_file "F";width
% 8.6042cm;height 8.9469cm;depth 0pt;original-width 7.9096in;original-height
% 8.2304in;cropleft "0";croptop "1";cropright "1";cropbottom "0";filename
% 'BNBT4diel.TIF';file-properties "XNPEU";}}

\begin{figure}[tbp]
\includegraphics[width=8.5 cm]{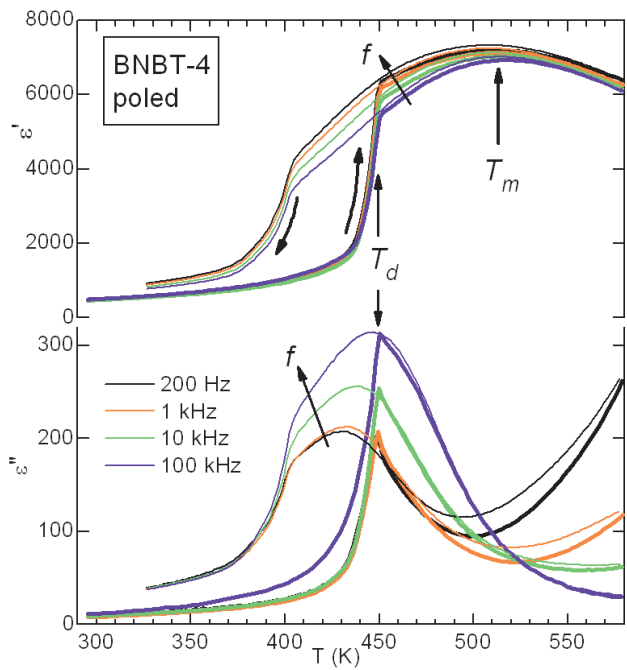}
\vspace{-3.5 cm}
\caption{Real and imaginary parts of the
permittivity $\protect\varepsilon $ of an initially poled sample of (NBT)$%
_{1-x}$(BT)$_{x}$ with $x=0.04$ (Color online).}
\label{fig BNBT4diel}
\end{figure}

Regarding the frequency dispersion of the dielectric maximum and the
presence of a secondary maximum with even stronger frequency dispersion near
$T_{d}$,\cite{Isu05b,SM01} Fig. \ref{fig BNBT4diel} shows the dielectric
permittivity of an initially poled sample with $x=0.04$ measured at
different frequencies. There is some frequency dispersion at $x\geq 0.04$,
although of reduced magnitude with respect to what found in typical relaxor
ferroelectrics, but we never found the secondary maximum near $T_{d}$; we
only find a neat step during heating in the poled state, with reduced
sharpness and temperature during cooling, or heating in the unpoled state.
The secondary maximum has been associated with an incommensurate modulation
between FE and AFE state,\cite{LYH04b,ZLL09} as found in the highly ordered
ferroelectric perovskite Pb(Yb$_{0.5}$Ta$_{0.5}$)O$_{3}$.\cite{YK93}
Incommensurate modulation of the R phase along the $c$ axis of the parent T
phase has indeed been found by neutron diffraction\cite{BKN06} and its
persistence after the disappearance of the T phase has been suggested to be
due to some type of ordering in the Na/Bi sublattice. Also TEM experiments
indicate a R phase modulated by twins,\cite{DTB08,DTB09} but at $T>T_{d}$.

Our dielectric and anelastic spectra do not show features necessarily
ascribable to incommensurate modulations; the anelastic losses remain high
also below $T_{2}$ and $T_{d}$, but this might also be due to generic domain
wall motion. Apparently, such walls may give rise to incommensurate
modulations, whose morphology likely depends on the sample quality. In this
respect, we recall that the dielectric hump that is presumed to accompany
incommensurate modulations is enhanced by introducing cation vacancies
through La doping\cite{YLH02b} or in solid solution with K$_{0.5}$Bi$_{0.5}$%
TiO$_{3}$\cite{SMK03,YCC05,XJS09} or BaTiO$_{3}$.\cite{GSM04} It is
therefore likely that the exact stoichiometry and degree of ordering of
cations determine the detailed structural distortions, their correlation
lengths and domain morphologies, and the scarce consistency between the
results of various experiments is not only due to differences in the
properties that are probed but also in the quality of the samples.

\subsection{Phase diagram}

\begin{figure}[tbp]
\includegraphics[width=8.5 cm]{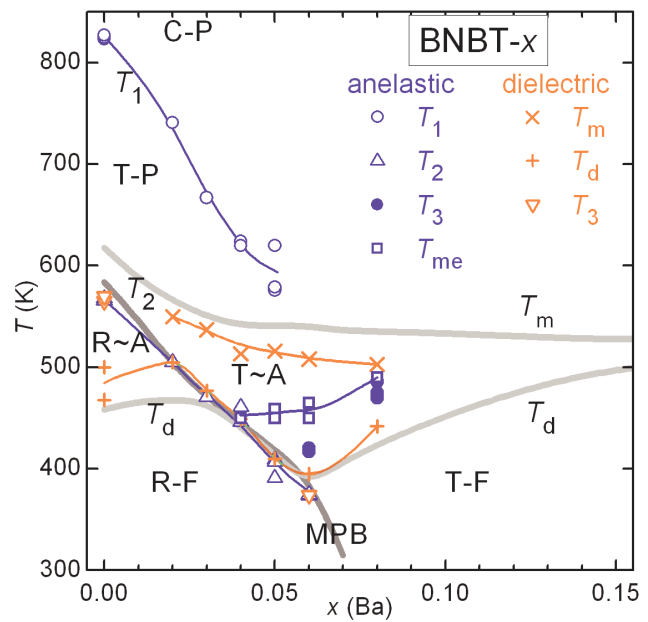}
\vspace{-4 cm}
\caption{Phase diagram of (NBT)$_{1-x}$%
(BT)$_{x}$deduced from the present anelastic and dielectric (10~kHz)
measurements. Grey lines: $T_{m}$, $T_{2}$, $T_{d}$ from Ref.
\onlinecite{HWN07}. C = cubic, T\ = tetragonal, R = rhombohedral, F
= ferroelectric, $\sim $A = almost antiferroelectric, P =
paraelectric (Color online).} \label{fig phasedia}
\end{figure}

Figure \ref{fig phasedia} shows the phase diagram of (NBT)$_{1-x}$(BT)$_{x}$%
deduced from the present anelastic and dielectric (10~kHz) measurements
during heating, together with that available in the literature\cite{HWN07}
(grey lines $T_{m}$, $T_{2}$, $T_{d}$). The latter is mainly based on
dielectric measurements and therefore does not include the border between T
and C phases. It is sometimes assumed that the paraelectric phase above $%
T_{m}$ is cubic, but this cannot be the case at low enough $x$, since the
neutron diffraction data show that NBT is tetragonal above $T_{m}$ and
becomes cubic only at $T_{1}=820$~K.

% \FRAME{ftbpFU}{8.6042cm}{8.3054cm}{0pt}{\Qcb{Phase diagram of (NBT)$_{1-x}$%
% (BT)$_{x}$deduced from the present anelastic and dielectric (10~kHz)
% measurements. Grey lines: $T_{m}$, $T_{2}$, $T_{d}$ from Ref.
% \onlinecite{HWN07}. C = cubic, T\ = tetragonal, R = rhombohedral, F =
% ferroelectric, $\sim $A = almost antiferroelectric, P = paraelectric (Color
% online).}}{\Qlb{fig phasedia}}{phasedia.tif}{\special{language "Scientific
% Word";type "GRAPHIC";maintain-aspect-ratio TRUE;display "USEDEF";valid_file
% "F";width 8.6042cm;height 8.3054cm;depth 0pt;original-width
% 3.8156in;original-height 3.6824in;cropleft "0";croptop "1";cropright
% "1";cropbottom "0";filename 'PhaseDia.tif';file-properties "XNPEU";}}

The anelastic spectra of Fig. \ref{fig anelvsx} allow us to draw a $%
T_{1}\left( x\right) $ line that drops rapidly toward $T_{m}$, roughly at
the composition of the MPB; similarly to the NBT-PbTiO$_{3}$ phase diagram.%
\cite{HP96} It is not clear whether $T_{1}$ merges with the $T_{m}\left( x\right)
$ line at $x>0.05$ or dissolves into an average cubic phase with local
tetragonal distortions. The three open circles at $x=0.05$ refer to three
different samples where the anomaly at $T_{1}$ is extremely shallow and
broadened. The region of the phase diagram defined by $x>0.05$ and $%
T_{d}<T<T_{m}$ is of uncertain nature. In (NBT)$_{1-x}$(BT)$_{x}$ co-doped
with Zr, with a particularly evident relaxor-like frequency dispersion in $%
\varepsilon ^{\prime }$ near $T_{d}$, such a region has been defined as a
frustrated region with coexistence of\ "multiple domain states".\cite{SSO01}
The fading away of the anelastic anomaly at $T_{1}$ with increasing $x$
suggests that for $x>0.05$ and $T>$ $T_{d}$ the phase is pseudocubic with
local tetragonal distortions whose correlation length must be very small.

The temperature $T_{m}$ of the broad maximum of $\varepsilon ^{\prime
}\left( T\right) $ is somewhat lower than in Ref. \onlinecite{HWN07}, while
the temperature $T_{me}$ of the maximum of the compliance is up to 60~K
lower than $T_{m}$. A similar phenomenon occurs in the relaxor
ferroelectrics Pb(Mg$_{1/3}$Nb$_{2/3}$)$_{1-x}$Ti$_{x}$O$_{3}$ and
(Pb/La)(Zr/Ti)O$_{3}$, where the dielectric maximum is at a temperature
higher than the maximum of the elastic compliance but also of the
susceptibility deduced from NMR relaxation\cite{105,116} (but, unlike
dielectric and anelastic spectroscopies, NMR probes also excitations with $%
q>0$, see Ref. \onlinecite{Vug06}). This has been explained as the result of
the different sensitivity of the three techniques to the polar on one hand
and non-polar rotational degrees of freedom on the other hand, all
exhibiting a relaxational dynamics which is faster for the rotational ones.%
\cite{116} Indeed, also the dielectric maximum in NBT-BT acquires a
dispersion in frequency with Ba doping. In NBT-BT, however, the R-type
rotational degrees of freedom undergo a phase transformation at $T_{2}$;
therefore, the elastic maximum at $T_{me}$ should involve other rotational
degrees of freedom, possibly also M-type rotations, which are coupled to the
polar modes and hence yield dielectric and elastic susceptibility maxima at
similar temperatures.

The $T_{2}\left( x\right) $ line separating the R from the T phases joins $%
T_{d}\left( x\right) $ very soon, at $x=0.03$ according to Hiruma \textit{et
al.}\cite{HWN07} and already at $x=0.02$ after our measurements. There is
some variability in the values of $T_{2}$ determined from the dielectric
curves, depending on the polarization state, but overall there is good
agreement with $T_{2}$ determined from the anelastic spectra on unpoled
samples. At $x\geq 0.06$ the $T_{d}\left( x\right) $ line starts rising,
while $T_{2}\left( x\right) $ continues to decrease, determining the MPB.

As noted in the previous paragraphs, the dielectric anomaly at $T_{2}$ may
be not very sharp, especially in unpoled samples, and the marked anomalies
measured at $0.02\leq x\leq 0.06$ may mainly reflect the ferroelectric
transition. The compliance, however, is almost unaffected by the
ferroelectric transition, and always presents a sharp step at $T_{2}$. This
means that the rotational instability of R type has a clear onset at a well
defined temperature $T_{2}$ also at compositions where (NBT)$_{1-x}$(BT)$%
_{x} $ displays properties typical of highly disordered or relaxor systems.
Notably, the sharp step in $s^{\prime }$ coexist with the broad maximum at $%
T_{me}$ and a T-C transition that is almost completely faded away at $x\geq
0.05$ (see Figs. \ref{fig anelvsx} and \ref{fig BNBT6}). It turns out,
therefore, that the polar modes are strongly affected by disorder,
characterized by short correlation lengths and undergo a diffuse or relaxor
transition; on the other hand, the R-type modes undergo a phase transition
which is well-behaved from the elastic point of view, presumably with a long
correlation length.

\subsection{Morphotropic Phase Boundary and possible monoclinic phase}

The composition-temperature phase diagrams of various solid solutions
between a perovskite which is rhombohedral at low temperature and one which
is tetrahedral, have an almost vertical line that separates the two phases
and is called morphotropic phase boundary (MPB). Among the most studied
ferroelectric systems of this type are PZT\cite{NC06} and PMN-PT.\cite{Dav07}
After decades of studies on PZT, it has been found that an intermediate
monoclinic phase exists in a narrow region at the MPB.\cite{GCP00} The
identification of such a monoclinic phase, where the direction of the
polarization may continuously vary between those in the R and T phases, was
possible from high-resolution synchrotron x-ray diffraction experiments and
careful analysis of the neutron diffraction data. The possibility for the
polarization to continuously rotate between the T and R phases through the
monoclinic one is still debated,\cite{FFN08,SSK07}\ but is generally
considered as the key factor that yields an easy change of polarization
under stress and therefore high electromechanical coupling at the MPB
compositions.\cite{FC00}

The solid solution NBT-BT at room temperature passes from R to T at the
composition $x\sim 0.06$, which is generally identified with the MPB, and
where the electromechanical coupling is maximum. Actually, there are few
studies aimed at defining the MPB line $T_{2}\left( x\right) $,\cite%
{TMS91,HWN07} which, to our knowledge, is defined by only two points at $%
x=0.06$ and 0.07. Other tetragonal perovskites that give rise to a MPB with
NBT are PbTiO$_{3}$\cite{HP96} and K$_{0.5}$Bi$_{0.5}$TiO$_{3}$, but in the
latter the composition of KBT at the border between R and T at room
temperature has been reported to range between 0.17 and 0.8.\cite{XJS09}

Besides better defining the MPB line, it would be interesting to investigate
whether also NBT-BT has an intermediate monoclinic phase that allows
continuous rotation of the polarization between the pseudocubic $\left[ 100%
\right] $ and $\left[ 111\right] $ directions. The identification of a
monoclinic phase by diffraction would certainly be more difficult for (NBT)$%
_{1-x}$(BT)$_{x}$ than for PZT, due to the larger disorder caused by the
coexistence of ions with different charges, Na$^{+}$, Bi$^{3+}$ and Ba$^{2+}$
in the A sublattice with respect to Zr$^{4+}$ and Ti$^{4+}$, which have the
same charge. The signature of the M\ phase should be more evident in the
elastic compliance versus temperature: in fact, the reciprocal Young's
modulus of PZT has a large peak in correspondence with the MPB,\cite{127}
that may be simply explained in terms of the usual Landau free energy
expansion with a coupling term linear in strain and quadratic in the
polarization, if the polarization may rotate in a monoclinic phase instead
of switching between R and T phases.\cite{127}

The case of (NBT)$_{1-x}$(BT)$_{x}$ is more complicated than PZT,
since there is no single transformation between T\ and M, but two
distinct processes: a R-T\ transformation at $T_{2}$ and a diffuse
transformation around $T_{me}$ and $T_{m}$. The first is well
behaved from the elastic point of view, with the expected sharp step
in the compliance at the R-type rotational instability.\cite{CS98}
If monoclinic distortions exist, they must be connected with the
diffuse process. The fact that it is the main feature in the
dielectric susceptibility indicates that it is due to polar modes,
while the strong peaked response in the elastic compliance recalls
the behavior of the monoclinic phase of PZT. At variance with PZT,
however, there should not be a monoclinic phase, but only local
monoclinic distortions with short correlation lengths, thereby
producing a diffuse response. This is in agreement with recent
NMR\cite{ASI08} and diffuse x-ray scattering
experiments,\cite{KBD03} both revealing additional displacements of
the Na/Bi ions away from the rhombohedral $\left\langle
111\right\rangle $ direction along $\left\langle 100\right\rangle $,
producing a local monoclinic symmetry. The existence of monoclinic
regions is also supported by a diffuse scattering
experiment,\cite{KBD03} from which it can be deduced that they are
shaped as platelets one or two cell thick and 20~nm wide.

\section{Conclusions}

We measured the complex dielectric permittivity and elastic compliance of
(NBT)$_{1-x}$(BT)$_{x}$ with $0\leq x\leq 0.08$. The anelastic spectra
contain anomalies at both the temperatures $T_{1}$ and $T_{2}$ of the
tetragonal-cubic and tetragonal-rhombohedral transitions, plus a broad
maximum at a temperature $T_{me}$, which is up to 60~K lower then $T_{m}$ of
the the diffuse dielectric maximum. The $T_{1}\left( x\right) $ line in the $%
x-T$ phase diagram drops rapidly with $x$ and can be followed up to $x\simeq
0.05$, where the transition becomes so diffuse and weak to be hardly
recognizable. At higher Ba substitution, the high temperature structure
should be pseudocubic with short range tetragonal distortions. Instead, the
elastic anomaly at $T_{2}$ is extremely sharp in pure NBT and remains
distinct up to $x=0.06$, demonstrating that there is a well defined onset of
the R-type octahedral rotations producing the R structure, in spite of the
numerous indications of extremely diffuse transition and structural
heterogeneity. The dielectric permittivity in our samples does not have a
secondary maximum with frequency dispersion around the ferroelectric
transition at $T_{d}$, as often reported; instead, we observe steps both at $%
T_{d}$ and $T_{2}$ (which however coincide for $0.02\leq x\leq 0.05$), whose
exact temperature and sharpness depend on the sample state and polarization.
These features are common to samples prepared under different temperature
and time schedules, if precautions are taken in order to avoid Bi loss
during sintering; otherwise, the anelastic and dielectric spectra indicate
extremely blurred structural transformations and defective material,
especially above 800~K, where Na diffusion becomes possible through the
vacancy mechanism.

The anelastic and dielectric anomalies should be connected to the non-polar
rotational modes of the octahedra and polar cation displacement as follows.
The M-type rotations giving rise to the T\ phase becomes unstable at $T_{1}$%
, but the correlation length of such modes rapidly decreases with Ba
substitution, so that at $x>0.05$ no long range T order exists any more. The
R-type rotations giving rise to the R distortion are unstable at $T_{2}$,
and maintain long range correlations up to $x=0.06$, which is the Ba
composition of the morphotropic phase boundary. The diffuse dielectric
maximum at $T_{m}$ and elastic maximum at $T_{me}$ should be due to polar
modes, whose correlation length remains short unless an external poling
field is applied or aging of the order of several days at room temperature
is allowed.

With Ba substitution, the broad peak in the elastic compliance becomes
preponderant over the R-T\ transition and, by analogy with the behavior of
PZT with monoclinic structure at the MPB, the possibility is discussed that
on the local scale the polarization may have an intermediate direction
between R and T.

\subsection*{Acknowledgments}

The authors sincerely thank Mr. C. Capiani for the skillful preparation of
the samples, P.M. Latino for improvements in the electronics for the
anelastic experiments, A. Morbidini and F. Corvasce for improvements in the
sample holders for the anelastic experiments.
F.C. and F.C. also thank E.K.J. Salje for helpful discussions.
Financial support by CNR under
the Free Theme Research RSTL \textquotedblleft Ferroelectric and relaxor
materials\textquotedblright , Grant n. 250 is gratefully acknowledged.


\begin{references}

\bibitem{DTB08}
V. Dorcet, G. Trolliard and P. Boullay, Chem. Mater. {\bf 20}, 5061 (2008).

\bibitem{JT02}
G.O. Jones and P.A. Thomas, Acta Cryst. B {\bf 58}, 168 (2002).

\bibitem{ASI09}
I.P. Aleksandrova, A.A. Sukhovsky, Yu.N. Ivanov, Yu.E. Yablonskaya and S.B.
Vakhrushev, Ferroelectrics {\bf 378}, 16 (2009).

\bibitem{FLG09}
A.I. Fedoseev, S.G. Lushnikov, S.N. Gvasaliya, P.P. Syrnikov and S. Kojima,
Phys. Sol. State {\bf 51}, 1399 (2009).

\bibitem{STS95}
I.G. Siny, C.-S. Tu and V.H. Schmidt, Phys. Rev. B {\bf 51}, 5659 (1995).

\bibitem{Suc02}
J. Suchanicz, J. Mater. Sci. {\bf 37}, 489 (2002).

\bibitem{NB72}
A.S. Nowick and B.S. Berry, {\it Anelastic Relaxation in Crystalline Solids}.
(Academic Press, New York, 1972).

\bibitem{TSS94}
C.-S. Tu, I.G. Siny and V.H. Schmidt, Phys. Rev. B {\bf 49}, 11550 (1994).

\bibitem{OK06}
T. Oh and M.-H. Kim, Mater. Sci. Engin. B {\bf 132}, 239 (2006).

\bibitem{HNT08}
Y. Hiruma, H. Nagata and T. Takenaka, J. Appl. Phys. {\bf 104}, 124106 (2008).

\bibitem{VIK85}
S.B. Vakhrushev, V.A. Isupov, B.E. Kvyatkovsky, N.M. Okuneva, I.P. Pronin, G.A.
Smolensky and P.P. Syrnikov, Ferroelectrics {\bf 63}, 153 (1985).

\bibitem{Suc98}
J. Suchanicz, Mater. Sci. Engin. B {\bf 55}, 114 (1998).

\bibitem{BKN06}
A.M. Balagurov, E.Yu. Koroleva, A.A. Naberezhnov, V.P. Sakhnenko, B.N. Savenko,
N.V. Ter-Oganessian and S.B. Vakhrushev, Phase Trans. {\bf 79}, 163 (2006).

\bibitem{PKF04}
J. Petzelt, S. Kamba, J. F{\'a}bry, D. Noujni, V. Porokhonskyy, A. Pashkin, I.
Franke, K. Roleder, J. Suchanicz, R. Klein and G.E. Kugel, J. Phys.: Condens.
Matter {\bf 16}, 2719 (2004).

\bibitem{AIS06}
I.P. Aleksandrova, Yu.N. Ivanov, A.A. Sukhovskii and S.B. Vakhrushev, Phys.
Sol. State {\bf 48}, 1120 (2006).

\bibitem{DTB09}
V. Dorcet, G. Trolliard and P. Boullay, J. Magn. Magn. Mater. {\bf 321}, 1758-
1761 (2009).

\bibitem{LGS00}
S.G. Lushnikov, S.N. Gvasaliya, I.G. Siny, I.L. Sashin, V.H. Schmidt and Y.
Uesu, Science {\bf 116}, 41 (2000).

\bibitem{KBD03}
J. Kreisel, P. Bouvier, B. Dkhil, P.A. Thomas, A.M. Glazer, T.R. Welberry, B.
Chaabane and M. Mezouar, Phys. Rev. B {\bf 68}, 014113 (2003).

\bibitem{HPK06}
J. Hlinka, J. Petzelt, S. Kamba, D. Noujni and T. Ostapchuk, Phase Trans. {\bf
79}, 41 (2006).

\bibitem{TWL07}
H.Y. Tian, D.Y. Wang, D.M. Lin, J.T. Zeng, K.W. Kwok and H.L.W. Chan, Science
{\bf 142}, 10 (2007).

\bibitem{HNT06}
Y. Hiruma, H. Nagata and T. Takenaka, Jpn. J. Appl. Phys. {\bf 45}, 7409-7412
(2006).

\bibitem{Gla72}
A.M. Glazer, Acta Cryst.. B {\bf 28}, 3384 (1972).

\bibitem{NC87}
U.J. Nicholls and R.A. Cowley, J. Phys. C: Solid State Phys. {\bf 20}, 3417-
3437 (1987).

\bibitem{CS98}
M.A. Carpenter and E.H.K. Salje, Eur. J. Mineral. {\bf 10}, 693 (1998).

\bibitem{Suc01}
J. Suchanicz, J. Phys. C: Solid State Phys. {\bf 62}, 1271 (2001).

\bibitem{QSS05}
Y. Qu, D. Shan and J. Song, Mater. Sci. Engin. A {\bf 121}, 148 (2005).

\bibitem{ASI08}
I.P. Aleksandrova,, A.A. Sukhovsky, Yu.N. Ivanov, Yu.E. Yablonskaya and S.B.
Vakhrushev, Phys. Sol. State {\bf 50}, 496 (2008).

\bibitem{SC80}
N. Setter and L.E. Cross, J. Appl. Phys. {\bf 51}, 4356 (1980).

\bibitem{PCK94}
S.-E.Park, S.-J.Chung, I.-T.Kim, K.S. Hong, J. Am. Ceram. Soc. {\bf 77}, 2641 (1994).

\bibitem{LYH01}
J.-K. Lee, J.Y. Yi and K.-S. Hong, Jpn. J. Appl. Phys. {\bf 40}, 6003
(2001).

\bibitem{LHK02}
J.-K. Lee, K.S. Hong, C.K. Kim and S.-E. Park, J. Appl. Phys. {\bf 91}, 4538 (2002).

\bibitem{WR05}
D.I. Woodward and I.M. Reaney, Acta Cryst. B {\bf 61}, 387 (2005).

\bibitem{HNT09}
Y. Hiruma, H. Nagata and T. Takenaka, J. Appl. Phys. {\bf 105}, 084112 (2009).

\bibitem{HWN07}
Y. Hiruma, Y. Watanabe, H. Nagata and T. Takenaka, Key Engineering Materials
{\bf 350}, 93 (2007).

\bibitem{Isu05b}
V.A. Isupov, Ferroelectrics {\bf 315}, 123 (2005).

\bibitem{SM01}
S. Said and J.-P. Mercurio, J. Electroceram. {\bf 21}, 1333 (2001).

\bibitem{GSM04}
J.-R. Gomah-Pettry, S. Sa{\"{\i}}d, P. Marchet and J.-P. Mercurio, J. Eur.
Ceram. Soc. {\bf 24}, 1165 (2004).

\bibitem{LYH04b}
J.-K. Lee, J.Y. Yi and K.S. Hong, J. Appl. Phys. {\bf 96}, 1174 (2004).

\bibitem{ZLL09}
C. Zhou, X. Liu, W. Li, C. Yuan and G. Chen, J. Mater. Sci. {\bf 44}, 3833
(2009).

\bibitem{YK93}
N. Yasuda and J. Konda, Appl. Phys. Lett. {\bf 62}, 535 (1993).

\bibitem{YLH02b}
J.Y. Yi, J.-K. Lee and K.S. Hong, Ferroelectrics {\bf 270}, 203 (2002).

\bibitem{SMK03}
J. Suchanicz, J.P. Mercurio, K. Konieczny and T.V. Kruzina, Ferroelectrics {\bf
290}, 161 (2003).

\bibitem{YCC05}
X. Yi, H. Chen, W. Cao, M. Zhao, D. Yang, G. Ma, C. Yang and J. Han, J. Cryst.
Growth {\bf 281}, 364 (2005).

\bibitem{XJS09}
H. Xie, L. Jin, D. Shen, X. Wang and G. Shen, J. Cryst. Growth {\bf 311}, 3626 (2009).

\bibitem{HP96}
K.S. Hong and S.-E. Park, J. Appl. Phys. {\bf 79}, 388 (1996).

\bibitem{SSO01}
S.A. Sheets, A.N. Soukhojak, N. Ohashi and Y.-M. Chiang, J. Appl. Phys. {\bf
90}, 5287 (2001).

\bibitem{105}
F. Cordero, F. Craciun and P. Verardi, Ferroelectrics {\bf 290}, 141 (2003).

\bibitem{116}
F. Cordero, M. Corti, F. Craciun, C. Galassi, D. Piazza and F. Tabak, Phys.
Rev. B {\bf 71}, 094112 (2005).

\bibitem{Vug06}
B.E. Vugmeister, Phys. Rev. B {\bf 73}, 174117 (2006).

\bibitem{NC06}
B. Noheda and D.E. Cox, Phase Transitions {\bf 79}, 5 (2006).

\bibitem{Dav07}
M. Davis, J. Electroceram. {\bf 19}, 23 (2007).

\bibitem{GCP00}
R. Guo, L.E. Cross, S-E. Park, B. Noheda, D.E. Cox and G. Shirane, Phys. Rev.
Lett. {\bf 84}, 5423 (2000).

\bibitem{FFN08}
J. Frantti, Y. Fujioka and R.M. Nieminen, J. Phys.: Condens. Matter {\bf 20},
472203 (2008).

\bibitem{SSK07}
E.P. Smirnova, A.V. Sotnikov, O.E. Kvyatkovskii, M. Weihnacht and V.V. Lemanov,
J. Appl. Phys. {\bf 101}, 084117 (2007).

\bibitem{FC00}
H. Fu and R.E. Cohen, Nature {\bf 403}, 281 (2000).

\bibitem{TMS91}
T. Takenaka, K. Maruyama and K. Sakata, Jpn. J. Appl. Phys. {\bf 30}, 2236
(1991).

\bibitem{127}
F. Cordero, F. Craciun and C. Galassi, Phys. Rev. Lett. {\bf 98}, 255701 (2007).

\end{references}
\end{document}